\documentstyle[sprocl]{article}

\input epsf

\newcommand{\nd}[1]{/\hspace{-0.6em} #1}
\def\lsim{\mathrel{\rlap {\raise.5ex\hbox{$ < $}}
{\lower.5ex\hbox{$\sim$}}}}

\def\be{\begin{equation}}
\def\ee{\end{equation}}
\def\bea{\begin{eqnarray}}
\def\eea{\end{eqnarray}}
\def\gappeq{\mathrel{\rlap {\raise.5ex\hbox{$>$}}
{\lower.5ex\hbox{$\sim$}}}}
 
\def\lappeq{\mathrel{\rlap{\raise.5ex\hbox{$<$}}
{\lower.5ex\hbox{$\sim$}}}}

\begin{document}
\title{NON-TRIVIAL INFRARED STRUCTURE 
IN (2+1)-DIMENSIONAL QUANTUM ELECTRODYNAMICS
AND NON-FERMI LIQUID BEHAVIOUR}

\author{ N.E. MAVROMATOS}

\address{P.P.A.R.C. Advanced Fellow, University of Oxford, 
Department of Physics, Theoretical Physics, 1 Keble Road,
Oxford OX1 3NP, United Kingdom}


\maketitle\abstracts{I review recent work on the infrared structure 
of ($2+1$)-dimensional Abelian Gauge Theories
and their application to condensed matter physics.
In particular, within a large-$N$ Schwinger-Dyson treatment, 
and including an {\it infrared momentum cut-off},  
I demonstrate the existence of a non-trivial
infrared fixed point of the Renormalization Group.
I connect this property to non-fermi liquid 
low-energy behaviour, 
and I attempt to draw some conclusions about 
the possible application of this approach 
to an understanding of the normal and superconducting 
phases of
planar high-$T_C$ superconducting Cuprates.}

\section{Introduction}

Low-dimensional ($(2+1)$-dimensional) gauge theories
have received great attention in recent years,
due to their possible connection
with the theory of planar high-temperature 
superconducting cuprates. 
Originally, there was great interest in 
gauge theories with parity and time reversal violation,
as a result of suggestions that high-temperature 
superconductivity might be due to deviations 
from normal Bose or Fermi statistics of the pertinent 
excitations (`anyonic superconductivity')~\cite{Laughlin,FHL}.
However, at present there
is no experimental evidence for 
such a Parity violation. 

In ref. \cite{dm,dmstat} 
a proposal was made for a simple
gauge-theory model which 
exhibits two-dimensional superconductivity without
parity violation. 
The starting point is the condensed matter $(2+1)$-dimensional
$t-j$ model, formulated on a bi-partite lattice, 
which we regard as a candidate description of 
the physics 
of the planar high-$T_c$ superconducting cuprates. 
The model includes both nearest and next-to-nearest neighbor 
interactions. The theory 
is originally formulated in terms of electron operators 
$C_{\alpha,i}$, with $i$ a lattice site index, and $\alpha=1,2$ 
a spin $SU(2)$ index. One then implements the {\it spin-charge} 
separation ansatz in its `slave-fermion' form~\cite{ioffe,dm,dmstat}
\begin{equation}
      C_{\alpha, i}=\psi^i z_\alpha^i 
\label{spincharge}
\end{equation}
to construct hole degrees of freedom, $\psi_i$,
to be termed holons, 
which are spinless Grassmann fields on the lattice, and spinon degrees of 
freedom, $z_\alpha$, which are $CP^1$ fields, obeying 
bosonic statistics. 
Notice that the ansatz (\ref{spincharge}) has a {\it hidden} 
$U(1)$ Abelian gauge symmetry~\footnote{Actually, as recently 
shown~\cite{fm}, one can find generalizations of the 
slave-fermion ansatz (\ref{spincharge}) admitting the full spin 
symmetry $SU(2) \otimes U_S(1)$, where $U_S(1)$
is a statistics changing group.
In the phase where the fermion mass 
is generated dynamically by the $U_S(1)$ interactions
the $SU(2)$ group is broken down to an Abelian subgroup.   
For our 
purposes here we shall concentrate only on the Abelian
case.} between the $\psi_i$ and $z$
constituents, under which the original $t-j$ model, 
formulated in terms of $C$,$C^\dagger$ operators
is {\it trivially} invariant~\cite{dm}. From 
arguments in a large-spin $S$ analysis
of the antiferromagnet~\cite{shankarorig,dm}, where 
intersublattice hopping of holons is assumed suppressed,  
one can assume 
{\it two kinds of holons}, one for each sublattice, i.e. 
$\psi_i$ in (\ref{spincharge}) carry a `sublattice' index.
Such a situation corresponds to a dominance of the 
next-to-nearest neighbor interactions of the $t-j$ model~\cite{dmstat}. 
\paragraph{}
The effective continuum lagrangian, describing the
physically-relevant degrees of freedom that lie
close to a single point on the fermi surface - which have 
been argued in ref. \cite{am} to be the most relevant 
interactions for our purposes-
can, then, be shown to
acquire the following low-energy form after $z$ integration:
\begin{equation} 
{\cal L}_{eff} = -\frac{1}{4g_{st}^2} f_{\mu\nu}^2(a)
+
\sum _{i=1}^{N}{\overline \psi}^i (x) (i\nd{\partial}  + \nd{a}\tau_3
+ \frac{e}{c} \nd{A})\psi ^i (x) 
\label{efflagr}
\end{equation}
where $A_\mu$ is the (external) electromganetic field,
$a_\mu$ is the statistical Abelian gauge field, associated
with the $U(1)$ symmetry of the ansatz (\ref{spincharge}), 
and $g_{st}^2$ is the dimensionful coupling (with dimensions of mass)
of this interaction, not to be confused with the 
four-dimensional electric charge. The model is characterized 
by a dynamically-generated scale $\alpha$, which 
enters via~\cite{app} 
$\alpha = g_{st}^2N/8$ (see the following section); $\alpha$ 
is to be identified with the ultraviolet 
cut-off $\Lambda$ of the lattice model, for reasons 
explained in ref. \cite{am}, to be mentioned briefly below. 
The three-dimensional character
of the model (\ref{efflagr}) is attributed to the planar 
character of the statistical model.

The fields $\psi$  
are doublets with respect to the sublattice
degree of freedom;
the $\tau _3$ structure,
which acts in this sublattice space,
indicates the opposite spin of the antiferromagnetic
(bi-partite) lattice structure of the underlying lattice~\cite{dm}.
The
`flavour number'
$N$, on the other hand,  
represents `internal degrees' of freedom, associated
with the orientation of the momentum vectors of the
quasiparticle excitations~\cite{gallavoti}
in expansions about a certain
point of a finite-size fermi surface, which is divided 
into cells. 
The continuum theory is obtained by linearizing about points
on the surface, the linearization being done by the introduction
of quasiparticles as defined in ref. \cite{gallavoti}.
The concept of the quasiparticles is essential in yielding the correct
scaling properties to be used in the renormalization
group approach~\cite{gallavoti,am}. 
It can be shown~\cite{am,shankar} 
that in the infrared regime we are interested in, 
$N \rightarrow N(\Lambda ) >> 1$,   
where
$\Lambda$ is the ultraviolet cut-off in
momenta measured above the fermi surface.

Notice that the fermion fields
$\psi $ in (\ref{efflagr})
have {\it zero} bare mass; this is the result 
of the way the continuum limit has been taken in the 
lattice model of ref.~\cite{dmstat}. 
Superconductivity 
is based on {\it dynamical mass} generation for the fermion
fields~\cite{dm}, which, as a result of the {\it even}
number of fermion species
due to the bi-partite lattice structure, 
is {\it parity conserving}~\cite{app}. 
Indeed, the mass generation  
provides a mechanism 
for 
the spontaneous breaking of the electromagnetic $U(1)$
symmetry, upon coupling the model to external 
electromagnetic potentials~\cite{dm}.

We now notice that  
the sublattice structure is irrelevant
for a study of the normal phase of the model, 
i.e. when no fermion mass gap is generated. {}From now on, therefore,
we concentrate on a single sublattice, ignoring
the $\tau_3$ `colour' (sublattice) structure of the fermions.
{}From this point of view,
the statistical gauge interaction in (\ref{efflagr}) plays
exactly the r\^ole
of the fermion-gauge interaction
of $QED_3$.

Recently~\cite{am} we have embarked on an analysis of the 
properties of $QED_3$ in its {\it normal phase}.
Our point was to study the {\it infrared} structure 
of the theory. We have provided evidence for 
the {\it revelance} of the fermion-gauge interaction
in a renormalization-group sense, which lead to 
a non-trivial
infrared fixed point.
According to recent arguments~\cite{shankar,am},
this 
leads to 
a departure
from Fermi liquid behaviour~\cite{shankar,am}.  
In this talk I shall review briefly the situation
that leads to this non-trivial infrared structure.

\section{Schwinger-Dyson Approach and Non-Trivial 
Infrared Structure of QED$_3$}

The structure of the model (\ref{efflagr}) 
can be studied by resumming $1/N$ terms in the 
so-called Schwinger-Dyson (SD) treatment. 
The method consists of evaluating 
the complete propagator $S_F$ for the (continuum) fermion fields $\psi (x)$: 
\begin{equation}
    S_F^{-1} = A(p)\nd{p} + B (p) 
\label{sdequation}
\end{equation}
where $A(p)$ is the wave function renormalization, 
and $B(p)$ is the gap function. The physical `dynamically generated'
fermion mass $\Sigma(p)$ is then defined as 
$   \Sigma(p)=\frac{B(p)}{A(p)}$. 
In the normal phase of the model $B(p)=0$. 

In the one-loop resummed $1/N$ limit,
the SD equations 
read~\cite{app} 
\begin{equation}
A(p)=1 - \frac{\alpha}{\pi^2N}\frac{1}{p^3} \int 
_0^\infty dk \frac{kA(k)G(p^2,k^2)}{k^2A(k)^2 + B(k)^2}I(p,k),
\label{one}
\end{equation}
where 
\begin{eqnarray}
&~&I(p,k) \equiv 
 \alpha ^2 ln\frac{p+k+\alpha}{|p-k|+\alpha} - 
\alpha (p+k - |p-k|)+ 2pk - \nonumber \\
&~&\frac{1}{\alpha}|p^2-k^2|
(p+k-|p-k|)-
\frac{1}{\alpha ^2}(p^2-k^2)^2\{ln\frac{p+k+\alpha}{|p-k|+\alpha}
-ln\frac{p+k}{|p-k|}\},  
\label{kernel}
\end{eqnarray} 
and 
\begin{equation}
B(p)=\frac{\alpha}{\pi^2 N} \frac{1}{p}\int_0^\infty dk
\frac{kB(k)G(p^2,k^2)}{k^2A(k)^2 + B(k)^2}\{4ln\frac{p+k+\alpha}
{|p-k|+\alpha}\} 
\label{oneb}
\end{equation}
where the Landau gauge has been assumed, and 
$\alpha \equiv \frac{g_{st}^2 N}{8}$ is the dynamically-generated
scale of the theory which remains finite as $N \rightarrow \infty$. 
$\Gamma _\mu = \gamma _\mu G(p^2,k^2)$ is a vertex function
whose precise form is dictated by gauge invariance, and in 
particular by a self-consistent solution 
of the so-called Ward-Takahashi identities~\cite{app,kondo,pen}.  
For our purposes we adopt the Pennington-Webb ansatz~\cite{pen,kondo}:
\begin{equation}
G(p^2,k^2) = A(k) 
\label{three}
\end{equation}
where chiral symmetry breaking occurs for
arbitrarily large $N$ \cite{pis}. 
The integrals in (\ref{one}) and (\ref{oneb}) 
are effectively cut-off
at $\alpha$, due to a sharp decay of the integrands
above this scale~\cite{app}. Hence $\alpha$ may be 
considered as an effective UV cut-off, to be identified 
with $\Lambda$ in the context of our statistical model.
Moreover, for reasons that will become 
clear below we also introduce an infrared cut-off
$\epsilon$.

Using the ansatz (\ref{three}), one
can then analyze the SD equations,
in the various regimes of momenta,
in terms of a running
coupling obtained from 
substituting the solution for $A(p)$ 
into  equation (\ref{oneb})  for the gap~\cite{higashijima,kondo,am}:
\begin{equation}
g_R(p, \epsilon ) = \frac{g_0}{A(p, \epsilon )}
\label{renorm}
\end{equation}
where $g_0 =8/\pi ^2 N$, 
$N$ is the number of fermion
flavours, and $\epsilon$ is an infrared cutoff.
The definition 
(\ref{renorm}) of the running 
coupling is also justified 
within the conventional 
framework of Gell-Mann-Low 
renormalization of the gauge model. 

The analysis of ref. \cite{am}
improved on earlier analytic approximations
for the kernel $I(p,k)$  
suggested in ref. \cite{kondo}. 
The existence of a non-trivial infrared fixed point 
was one of the the two main conclusions 
reached by our analysis.
The second important 
result,
pertains to a significant
slowing down of the rate of decrease of $g_R$ with increasing $p$
at intermediate scales of momenta,
relative to that shown in the earlier treatments
of $QED_3$ in ref. \cite{kondo}. The situation is 
shown in figure 1.

\begin{figure}[htb]
\epsfxsize=3.2in
\centerline{\epsffile{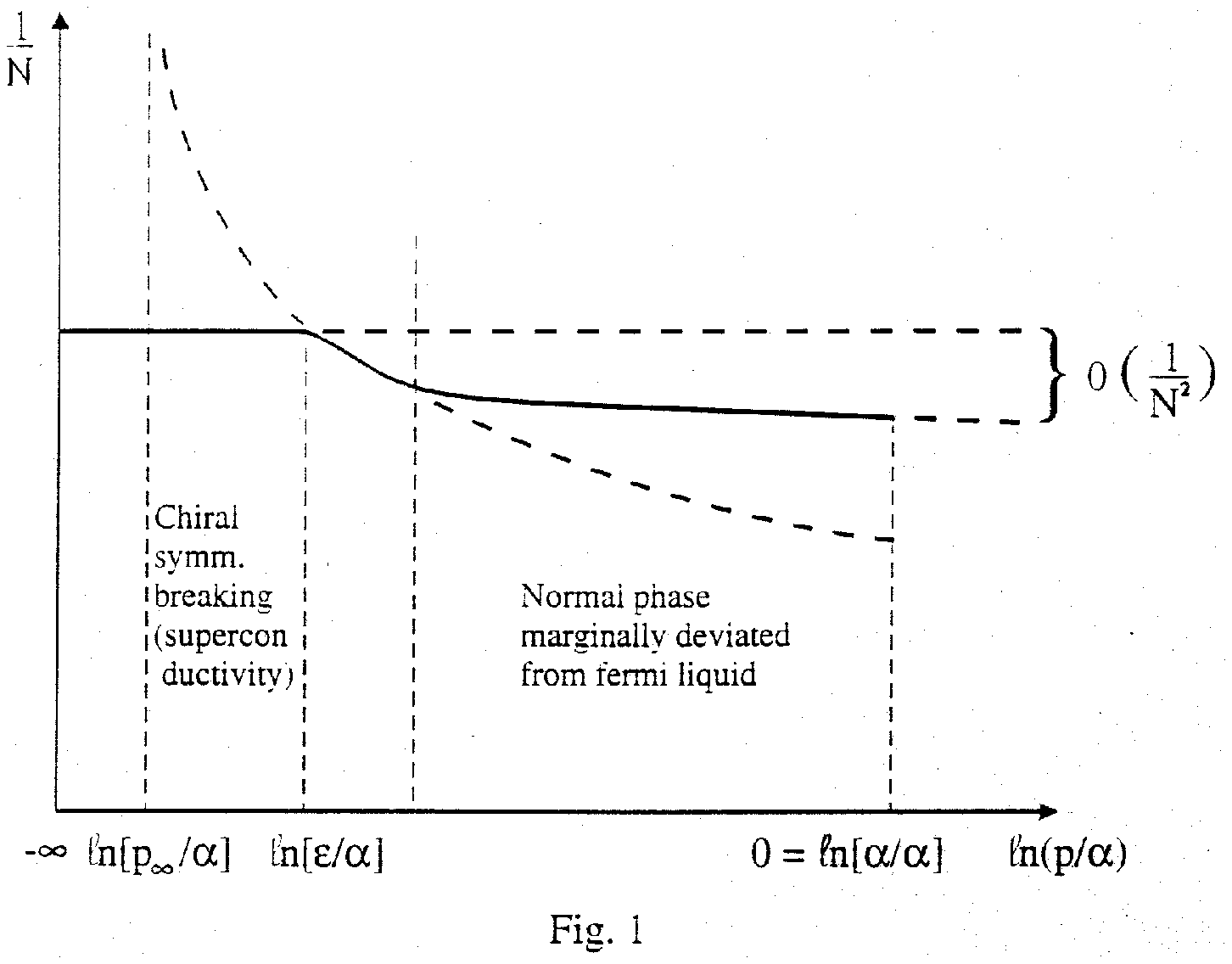}}
\caption{Running flavour number in $QED_3$ versus 
a momentum scale $p$, within a 
resummed-$1/N$ Schwinger-Dyson treatment, in the presence of  
an infrared cut-off $\epsilon$. The dashed lines pertain
to earlier treatments. The running is slow, and there exists 
a non-trivial infrared (IR) fixed point. This kind of behaviour 
is argued to be responsible for (marginal) deviations 
from the fermi-liquid (trivial IR fixed point) theory.
}
\label{fig1}
\end{figure}

In the analysis of ref. \cite{am} 
we have implemented the infrared cut-off 
in two ways: 
\begin{itemize} 
\item[(i)] as a simple momentum-space cut-off
${\hat \epsilon}$, in which case the wave-function 
renormalization becomes 
\begin{equation} 
A(p, \epsilon)= 1 -\frac{\alpha}{\pi^2 N}\frac{1}{p^3}
\int _{{\hat \epsilon}} ^\infty \frac{dk}{k} I(p,k)
\label{improvedeps}
 \end{equation}
which has been evaluated numerically in ref. \cite{am}. 

\item[(ii)] in a covariant way, 
by keeping the limits of 
integration from $0$ to $\alpha$, and interpreting the 
mass function $B/A$ in (\ref{one}) as a 
(covariant) infrared cut-off in the case of 
no dynamical mass generation. The expression for $A$ 
then reads
\begin{equation} 
A(p, \delta )=1 - \frac{\alpha}{\pi^2N}\frac{1}{p^3} \int 
_0^\alpha dk \frac{k}{k^2+ \delta ^2}I(p,k),
\label{onec}
\end{equation}

This way of introducing the infrared (IR) 
cut-off makes some contact with
the finite temperature case ~\cite{am,aklein}, where the 
plasmon mass was interpreted as an effective infrared cut-off.
The above similarity is, however, only indicative. Whether the 
situation described here carries over intact to the finite-
temperature regime is at present only an expectation.
These are issues that are left open for future investigations.

\end{itemize} 

It interesting to note that the running
shown in fig. 1 is 
characteristic of the presence of a {\em finite} 
infrared cut-off. Removal of  
$\delta$ via $\delta 
\rightarrow 0$ in a smooth manner does not seem to
be possible~\cite{am}. From 
a physical point of view ~\cite{am}, where the 
(covariant) infrared cut-off $\delta$ is conjectured to be 
associated with temperature in certain 
condensed-matter systems whose physics the above 
model is supposed to simulate, this would imply 
that the above non-trivial structure is an 
exclusive feature of the finite-temperature
field theory~\footnote{It worths noticing, however, 
that a recent study~\cite{kondo2},
which improved on the approximations of ref. \cite{am}
by solving the SD equations in the {\it non-local} gauge,  
has shown that removal of the ${\hat \epsilon}$ cut-off, in the 
sense of ${\hat \epsilon} \rightarrow 0$, may be possible,
whilst it has 
confirmed our results for the $\delta$ cut-off.}.

\section{Outlook} 

Numerical solution of the SD equations for multiflavour $QED_3$ 
indicated~\cite{am} that the increase of the running coupling
is cut-off in the infrared, in the form of a non-trivial fixed 
point, and that there is also a significant slowing down 
(`walking' behaviour)
of the running of the coupling at intermediate scales,
as compared to the case of ref. \cite{kondo}.
Both features, 
have been  argued in ref. \cite{am} 
to be responsible for deviations from fermi-liquid 
behaviour.  This might have important 
physical consequences,  in case the model  
simulates correctly  the physics of the novel high-temperature
superconductors. It should be stressed that 
the above picture is valid only for three dimensions, and 
thus materials with {\it planar} structure.
We have also argued~\cite{am} that the above 
non-trivial low-energy structure is a consequence of a 
{\it finite} infrared cut-off. If one 
associates~\cite{am} the infrared cut-off
with temperature effects, this would imply 
that the above non-trivial infrared structure is an exclusive feature 
of the finite-temperature field theory. 

An important issue concerns
the connection of 
these results 
with the situation in the 
three-dimensional Thirring model.
In this conference we have seen 
evidence for the existence of a non-trivial 
ultraviolet fixed point structure 
of the Thirring model, coming from lattice simulations~\cite{hands}.
In view of the conjectured equivalence 
between the Ultraviolet structure of the Thirring 
model and the Infrared structure of $QED_3$~\cite{hands},
our results offer 
non-trivial support to the above conjecture. 
More work clearly needs to be done before definite conclusions
are reached, but we believe that the potential 
application of these results to realistic condensed-matter
systems makes such studies physically
relevant and worth pursuing.



\end{document}